\documentclass[prl,aps,twocolumn,showpacs,amsmath,amssymb]{revtex4}

\usepackage{epsf}

\begin{document}
\draft
\newcommand{\ve}[1]{\boldsymbol{#1}}

\title{Emerging magnetism and electronic phase separation at titanate interfaces}

\author{N.~Pavlenko$^{1,2}$, T.~Kopp$^{1}$, and J.~Mannhart$^{2}$}
\address{$^1$ EKM and Institut f\"ur Physik, Universit\"at Augsburg, 86135 Augsburg, Germany \\
$^2$ Max Planck Institute for Solid State Research, Heisenbergstr.1, Stuttgart, Germany}

\begin{abstract}
The emergence of magnetism in otherwise nonmagnetic compounds and its underlying mechanisms have become the subject of intense research. Here we demonstrate that the nonmagnetic oxygen vacancies are responsible for an unconventional magnetic state common for titanate interfaces and surfaces. Using an effective multiorbital modelling, we find that the presence of localized vacancies leads to an interplay of ferromagnetic order in the 
itinerant $t_{2g}$ band and complex magnetic oscillations
in the orbitally-reconstructed $e_g$-band, which can be tuned by gate fields at oxide interfaces.
The magnetic phase diagram includes highly fragmented regions of stable and phase-separated magnetic states forming beyond nonzero critical defect concentrations.
 \end{abstract}


\date{\today}

\maketitle

The possibility to induce a magnetic state at interfaces and surfaces of otherwise nonmagnetic titanates 
by introducing defects like oxygen vacancies is an open question that is in the focus of present research \cite{hwang,brinkman,luli,bert,kalisky,ariando,kalabukhov,berner,joshua,lee,liu,salluzzo,
han,jeschke,hergert,sawatzky}.
More recent  studies of this 
problem are related to the interfaces between the 
bulk insulators LaAlO$_3$ (LAO) and SrTiO$_3$ (STO)~\cite{hwang,millis,chen,michaeli,zhong,macdonald}, for which low-temperature torque magnetometry, scanning SQUID and magnetoresistence 
studies \cite{luli,bert,ariando,kalisky} provide evidence for an inhomogeneous electronic state with coexisting superconducting (diamagnetic), ferromagnetic/superparamagnetic
and paramagnetic mesoscale regions.     
Unlike carrier-induced ferromagnetism in diluted magnetic semiconductors, the oxygen defects are
intrinsically nonmagnetic which implies that the build-up
of magnetism should involve a magnetization of defects at an initial stage. 

New insight into the local electronic state of oxygen-reduced titanate layers has been recently gained from
DFT studies \cite{pavlenko,pavlenko2,pavlenko4} that
identified stable local magnetic moments of spin-polarized Ti $3d$-states in the proximity of oxygen defects, which induce quasi two-dimensional (2d) ferromagnetic order at the LAO/STO interface.
Although the DFT-studies 
associate the interface magnetism to the local nonstoichiometries, they 
frequently impose artificially ordered states by 
inevitably restricting the analysis to selected supercell configurations.
Such DFT studies are usually limited to the consideration of local electron states with high defect concentration \cite{pentcheva,tsymbal}.  

As a consequence, the build-up of the spin-polarization and inhomogeneous state at titanate interfaces
for different levels of defect concentration and its relation to
random defects like oxygen vacancies remains unresolved.
To understand the mechanism of the spin-polarization in the presence of a finite density of oxygen defects and for different mobile carrier densities, 
we consider a microscopic modelling of the 2d electron liquid at titanate interfaces in an external electrostatic field applied accross the gate STO-layer.  

The challenge that we face is to explore if a multiband model 
with randomly occupied defect band
allows for the formation of the robust magnetic state at the micro- and nanoscales---even though 
the defects are non-magnetic in the dilute limit.
In particular, is there a finite critical density of defects where magnetism sets in? 
DFT studies are not appropriate to assess whether such a critical density exists.

\begin{figure}[tbp]
\epsfxsize=8.5cm {\epsfclipon\epsffile{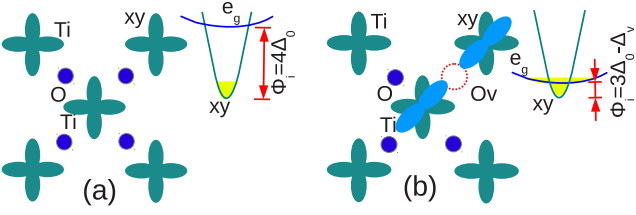}}
\caption{Scheme of vacancy-induced orbital reconstruction in 
the STO interface layer with a strong negative shift and partial electron 
occupation of the $e_g$ orbital (b) of a Ti$_2$O$_4$-plaquette. The initial
stoichiometric configuration with electron-occupied $3d_{xy}$ levels of Ti is shown in (a). The occupied electron states are yellow-colored.}
\label{fig1}
\end{figure}

For clarity we analyze a two-orbital model on an $N$-site square lattice with random occupation of the second orbital. The Hamiltonian below is to be taken as an effective Hamiltonian for the 2d electron system 
at the LAO/STO interface with the lowest Ti $t_{2g}$ band partially occupied, viz.\ the itinerant  Ti $3d_{xy}$ band (label $\alpha=1$), and with an oxygen-defect relevant Ti $e_g$ band (label $\alpha=2$). 
In contrast to stoichiometric SrTiO$_3$ with empty high-energy $e_g$ bands, 
the $3d$ orbital reconstruction due to the oxygen vacancies \cite{pavlenko2} modifies the local covalent bonding and shifts down 
the local $e_g$-level of the Ti in the proximity of vacancy, which leads to a random electronic occupation 
of the band $\alpha=2$ by vacancy-released $3d$-electrons (Fig.~\ref{fig1}):
\begin{eqnarray} \label{hor}
H_{OR}= (\varepsilon_d-\mu) \sum_{i=1\atop \sigma=\uparrow,\downarrow}^N n_{i,1,\sigma}
+\sum_{i=1\atop \sigma}^N (\varepsilon_d-\mu+\phi_{i\sigma}) n_{i,2,\sigma}
\end{eqnarray}
Here the electron number operators $n_{i,\alpha,\sigma}=c_{i,\alpha,\sigma}^{\dag} c_{i,\alpha,\sigma}$, where 
$c_{i,\alpha,\sigma}^{\dag}$ are electron creation operators, $\sigma$ is the electron spin index, $\mu$ is the chemical potential, and
$\varepsilon_d$ is the $3d_{xy}$ reference energy level of Ti.
The field $\phi_{i\sigma}=\sum_{l=1,4}(x_{il}\Delta_o-(1-x_{il})\Delta_v)$ depends on 
the local values of random variables $\{ x_{il}\}=0,1$ which correspond to the presence or absence
of oxygen atoms $(i,l)$ in the nearest oxygen configuration 
around the $i$-th Ti atom. 
In the stoichiometric heterostructure, the octahedral field splits the two considered $3d$ levels by a gap $4\Delta_o$ which is obtained
by setting all local variables $x_{il}=1$ in the     
field $\phi_{i\sigma}$.
In contrast, in the system with oxygen defects, recent DFT studies of LAO-STO \cite{pavlenko2} reveal an
electronic occupation of the Ti $e_g$ state due to oxygen vacancies in the positions $(i,l)$. The resulting negative shift $-\Delta_v$ of the $e_g$ energy level is described by setting  
the value of $x_{il}$ in (\ref{hor}) to zero. From the DFT studies, we
estimate the energies $\Delta_o\approx 0.7$eV and $\Delta_v\approx$1.2--2.4~eV.
Additional 
terms describe the local Coulomb repulsion and
kinetic electron inter- and intra-orbital exchange (cf.~\cite{fazekas})
\begin{eqnarray} \label{hc}
&& H_{C}=U\sum_{\alpha,i} n_{i,\alpha, \uparrow} n_{i,\alpha, \downarrow}+U'\sum_{i} n_{i,1} n_{i,2} \\
&& + J_H \sum_{i} (b_{i,\uparrow}b_{i,\downarrow}+h.c.)-2J_H \sum_{i} \vec{S}_{i1} \vec{S}_{i2} \nonumber \\
&& -\sum_{\alpha; \sigma; \langle i,j\rangle} t_{\alpha}
(c_{i,\alpha,\sigma}^{\dag} c_{j,\alpha,\sigma}+h.c.)\nonumber
\end{eqnarray}
Here $b_{i\sigma}=c_{i,1,\sigma}^{\dag}c_{i,2,\sigma}$ is an interband  operator;
$U=2$~eV is the Coulomb repulsion energy for $3d$ Ti electrons~\cite{breitschaft}, $J_H$ is the Hund coupling energy,
and $U'=U-2J_H$. The three-component band spin operator $S_{i\alpha}$ is defined
as $S_{i\alpha}=\left\{ c_{i,\alpha,\uparrow}^{\dag} c_{i,\alpha,\downarrow},
c_{i,\alpha,\downarrow}^{\dag} c_{i,\alpha,\uparrow}, \frac{1}{2} (n_{i,\alpha,\uparrow}-n_{i,\alpha,\downarrow}) \right \}$, and 
the electron orbital operators as $n_{i,\alpha}=\sum_{\sigma} n_{i,\alpha,\sigma}$.  
From the DFT band structure calculations, we obtain
the effective electron transfer energy $t_1=0.28$~eV of the lowest $3d_{xy}$ Ti band. The strong-localization character of the orbitally-reconstructed
$e_g$ band results from a much smaller value of $t_2$,
viz.\ $t_2/t_1=0.05$.
In the following analysis, the energy parameters are scaled 
by $t_1$. 

\begin{figure}[tbp]
\epsfxsize=8.cm {\epsfclipon\epsffile{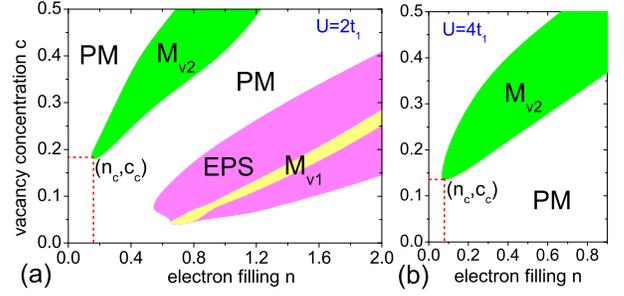}}
\caption{Typical diagram of magnetic phases predicted for LAO-STO interfaces. The phase diagram (a) is obtained for
$J_H/t_1=1.8$, $k_BT/t_1=0.03$, $\Delta_o/t_1=2.0$, 
$\Delta_v/t_1=6.0$, $U/t_1=2$. 
The red color marks the regimes with electronic phase separation (EPS) of paramagnetic and magnetic state $M_{v1}$.
(b) Detailed view of the phase $M_{v2}$ for $U/t_1=4$.}
\label{fig2}
\end{figure}

The interaction of electron charges with polar electrostatic fields 
shifts the electron energies $H_{FE}=-E_g\sum_{i,\alpha} n_{i\alpha}$
by the value $E_g=(de)V_g$, where $V_g$ is the external gate field \footnote{For a 100~\AA\,-thick gate layer, the typical values of the gate potential $V_g\sim 10^{6}$~V/cm 
induce maximal electron energy shifts $E_g$ of about $1.0$~eV.}, 
$d$ is the thickness of the gate layer and $e$ is the electron charge \cite{pavlenko_kopp,pavlenko_schwabl}. 

\begin{figure}[tbp]
\epsfxsize=8.6cm {\epsfclipon\epsffile{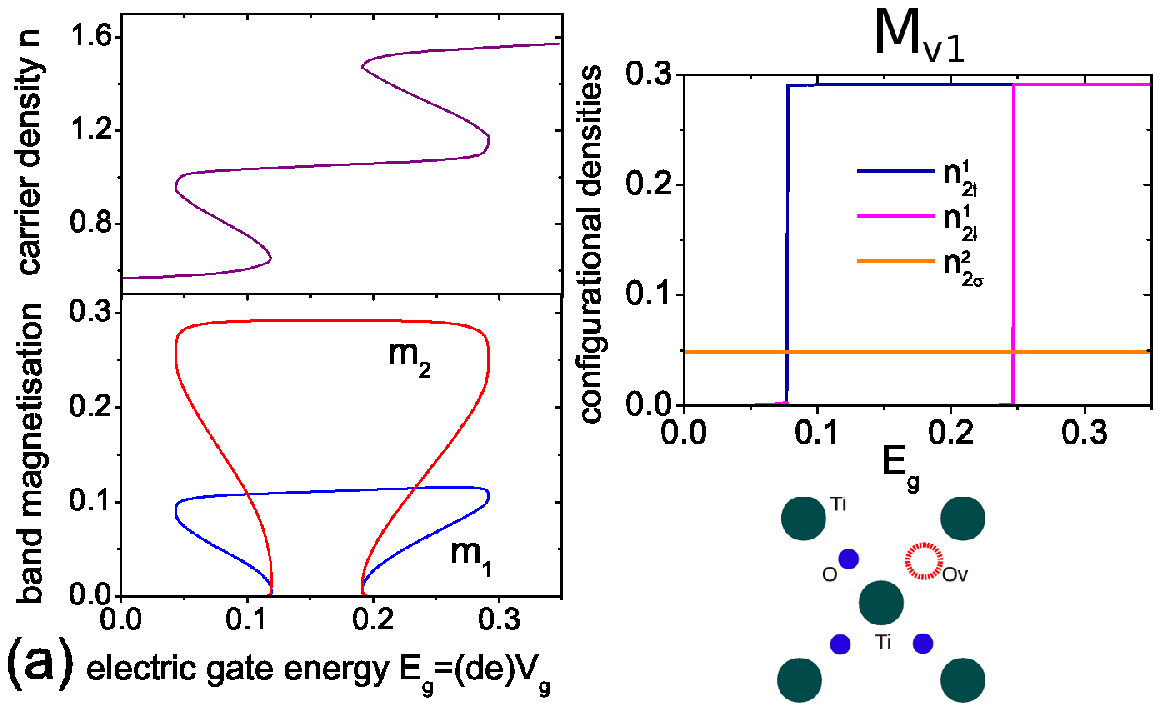}}
\epsfxsize=8.6cm {\epsfclipon\epsffile{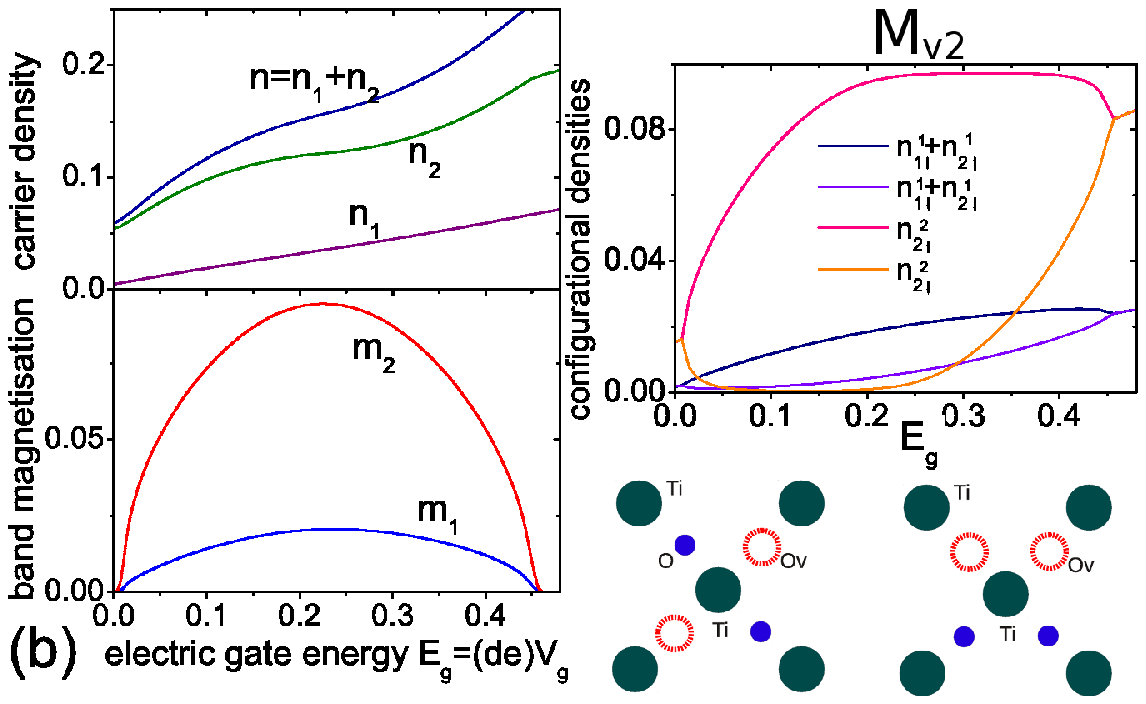}}
\caption{Electron carried density and band magnetisation versus electric-gate-induced energy $E_g$
(a) in the phase $M_{v1}$, with $\mu/t_1=-1.4$, $c=0.1$, $U=2t_1$ and (b) in the phase $M_{v2}$, with $\mu/t_1=-3.9$, $c=0.15$, $U=4t_1$. 
Here $J_H=1.8t_1$, $k_BT=0.03t_1$, $\Delta_o=2t_1$. 
The plots at the rhs show the variation of the configurationally resolved contributions to the band occupancies.
The superindices refer to the number of O-vacancies next to the central Ti-atom in the displayed clusters.
}
\label{fig3}
\end{figure}

\begin{figure}[tbp]
\epsfxsize=8.5cm {\epsfclipon\epsffile{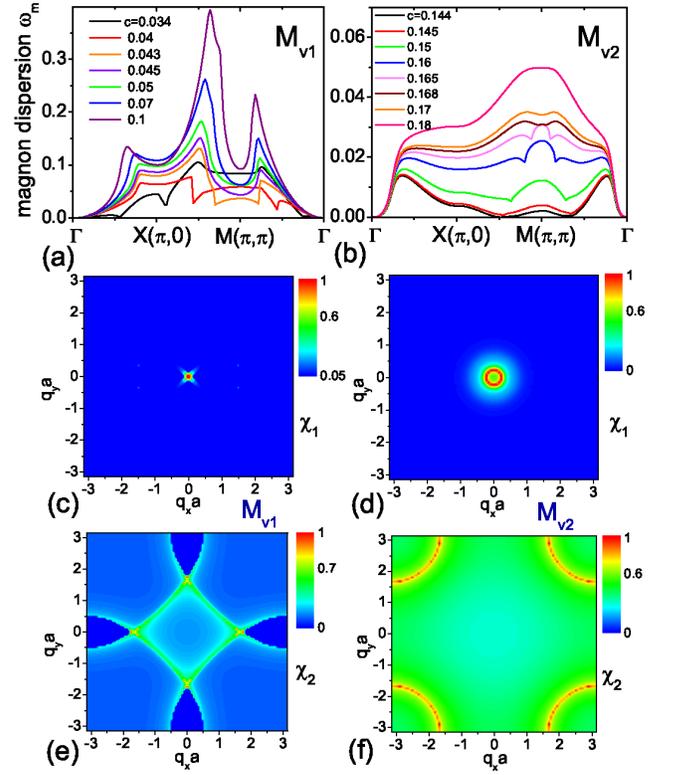}}
\caption{Magnon dispersion energies for different defect concentration $c$ (a) in the $M_{v1}$ state with $U=2t_1$, 
$\Delta_v=6t_1$ and (b) in the $M_{v2}$ state with $U=4t_1$, $\Delta_v=4t_1$.
Here $J_H/t_1=1.8$, $k_BT/t_1=0.03$, $\Delta_o/t_1=2.0$, $\Delta_v/t_1=4.0$.
The contours (c),(d), and (e),(f) display the imaginary part of the Lindhard functions $\chi_{\alpha}^{+-}(\vec{q},0)$ (scaled by their maximal values) for $\alpha=1$ and $\alpha=2$, respectively: left, in the magnetic state $M_{v1}$, and right, in $M_{v2}$.}
\label{fig4}
\end{figure} 

To analyze the thermodynamic state of the model (\ref{hor})--(\ref{hc}) in a weak and moderate Coulomb-coupling regime, we employ a generalized molecular-field approach, which contains, besides the mean-field type decoupling of the correlations terms in (\ref{hc}) and thermodynamical averaging, an
additional configuration averaging  with respect to the local random
configurational fields $\phi_i$.
We assume independence of each local configuration in the $i$-th unit cell on the local configurations
of the surrounding cells, in the spirit of the virtual-crystal approach (VCA)~\cite{elliot}. The resulting configurationally averaged grand canonical thermodynamic
potential is given by 
\begin{eqnarray*}
\frac{\Omega}{N}\!=\! -\xi_0-k_BT(1-c)^4 \!\!\!\! \sum_{j=0\atop \alpha=1,2; \sigma}^4 \sum_{\vec{k}} p_j \ln (1+e^ {-\lambda_{\alpha\sigma}^j(\vec{k})/k_BT})
\end{eqnarray*}
where the configurational weight $p_j=p_j^0 \left(\frac{c}{1-c} \right)^j$ depends on the average concentration (probability) of the oxygen defects $c=1-x$ 
determined 
through $x=\frac{1}{4N} \sum_{i,l} \langle x_{il} \rangle$; 
the vector $\{p_j^0\}=\{1,4,6,4,1\}$ contains the statistical contributions of random vacancy configurations of each TiO$_4$-plaquette 
with $j=0,\ldots,4$ vacancies near the Ti-atom. 
The orbital- and spin-projected band energies are given by $\lambda_{\alpha\sigma}^j(\vec{k})=\Pi_{\sigma}(\vec{k})+\phi_{\sigma}^j/2-(-1)^{\alpha}{s_{j\sigma}}(\vec{k})/{2}$, 
where 
$\Pi_{\sigma}(\vec{k})=\varepsilon_d-\varepsilon_g-\mu+\frac{3}{2}(\frac{U}{2}-J_H)n-{\sigma}\frac{U+J_H}{4}(m+t\eta_{\vec{k}})$ with the magnetization $m=\sum_{\alpha}m_{\alpha}$
and the kinetic parameter $t=\sum_{\alpha}t_{\alpha}$ with $\eta_{\vec{k}}=\cos k_x +\cos k_y$. The interband splitting $s_{j\sigma}(\vec{k})=\sqrt{(\Delta \varepsilon_{\sigma}(\vec{k})-\phi_{\sigma}^j)^2+4J_H^2|\langle b_{\sigma}\rangle|^2}$ depends on the interorbital gap $\Delta \varepsilon_{\sigma}(\vec{k})=(-\frac{U}{2}+3J_H)\delta-{\sigma}\frac{U-J_H}{2}(m_1-m_2)+2(t_1-t_2)\eta_{\vec{k}}$,
the charge order parameter $\delta=\langle n_1\rangle - \langle n_2\rangle$, and orbital magnetization $m_{\alpha}=\langle n_{\alpha\uparrow}\rangle - \langle n_{\alpha\downarrow}\rangle$. For each unit cell, the local field vector $\{ \phi_{\sigma}^j \}=\phi_{i\sigma}(\{ x_{il}\})=\{ 4\Delta_o; 3\Delta_o-\Delta_v; 2(\Delta_o-\Delta_v); \Delta_o-3\Delta_v; 4\Delta_v\}$
represents possible values of $\phi_{i\sigma}$ in the local random vacancy-configuration environment with $j$ vacancies near each Ti atom. The quadratic term $\xi_0=\frac{U}{4}(\frac{1}{2}(n^2+\delta^2)-(m_1^2+m_2^2))+\frac{1}{4}(U-3J_H)(n^2-\delta^2)+J_H(\langle b_{\uparrow}^{\dag}\rangle \langle b_{\downarrow}\rangle+h.c.)$ contains the thermodynamically and configurationally averaged order parameters which can be self-consistently determined from the corresponding extremum conditions for $\Omega$.    

To explore the magnetic states, 
we investigate interface thermodynamic phases 
by minimization of $\Omega$ with respect to $m_{\alpha}$, $\langle b_{\sigma} \rangle$ and $\delta$. Fig.~\ref{fig2} shows a characteristic phase diagram ($n$, $c$). 
For nonzero defect concentration $c$, we find, besides the trivial paramagnetic
state corresponding to $m_{\alpha}=0$, two different magnetic phases indicated by $M_{vj}$ ($j=v1,v2$), where the indices $v$ and $j=1,2$ signify the physical nature of magnetic ordering induced by vacancies. 
The magnetic phases $M_{v1}$ and $M_{v2}$ are characterized by non-zero polarization in both, 
itinerant ($\alpha=1$)- and localized ($\alpha=2$)-bands. 
The magnetic regions are highly fragmented.
The phase $M_{v1}$ is bounded by the red-colored regions 
of electronic phase separation characterized by the separation of the electron 
liquid into lower-concentration paramagnetic and higher-concentration spin-polarized states. 

The contributions of different local vacancy configurations to the microscopic parameters like band filling $\langle n_{\sigma \alpha} \rangle$ and magnetization $m_{\alpha}$
are related to the random field vector $\left \{ \phi_{\sigma}^j \right \}$ and depend on the reconstruction parameter $\Delta_v$. 
Small values of $\Delta_v \le 3\Delta_o=6t_1$ correspond to a small and positive one-vacancy local field component  $\phi_{\sigma}^1=3\Delta_o-\Delta_v=2t_1$ and result in negligible contributions of one-vacancy-configurations to the occupancy $\langle n_2 \rangle$ of the localized band ($\alpha=2$), which becomes more significant only for large $n>1$. In this case, the 
pronounced negative shift of the localized ($\alpha=2$)-band level comes from the two-vacancy configurations
represented by the local field component $\phi_{\sigma}^2=2(\Delta_o-\Delta_v)$, which is close to the stripe- and cluster-type configurations with two vacancies near each Ti-atom. Furthermore, the less clusterized one-vacancy (Ti-O$_v$-Ti dimer) configurations become important 
as the reconstruction parameter is increased to $\Delta_v \ge 6t_1$. This increase leads to zero or negative value of the field $\phi_{\sigma}^1$ and in this way sets the energies of the itinerant ($\alpha=1$)- and localized ($\alpha=2$)-bands to the same level. As a result, the parameter $\Delta_v$  determines the configurational character of the electronic properties and describes the clusterization level of the oxygen vacancies which can be controlled by vacancy diffusion in external electric fields. 

We address now the properties of the magnetic states $M_{v1}$ and $M_{v2}$. Both $M_{v1}$ and $M_{v2}$ appear at finite critical defect concentrations and are induced by oxygen defects. 
A remarkable feature of the phase diagrams in Fig.~\ref{fig2} is the existence of the critical (cusp) point ($c_c$, $n_c$) terminating the vacancy-induced magnetic state $M_{v2}$.
The cusp implies the existence of a lowest nonzero vacancy 
concentration $c_c \approx 0.1$ that is necessary for the stabilization of the state $M_{v2}$. The decrease of the Coulomb repulsion $U$ from $U/t_1=4$ (Fig.~\ref{fig2}(b)) to $U/t_1=2$ (Fig.~\ref{fig2}(a)) increases the critical vacancy concentrations $c_c$ from $0.13$ to $\sim 0.2$.
The origin of the spin polarization in this low-concentration range is beyond the well-known intrinsic 
property of the Hubbard model to stabilize the ferromagnetic state upon an increase of $U$ \cite{hirsch,pavlenko_1dcpa}. 
To understand the origin of the magnetism, it is instructive to consider the configuration- and spin-projected  band occupancies $n_{\alpha\sigma}^j=p_j\sum_{\vec{k}} [1+\exp(\lambda_{\alpha \sigma}^j(\vec{k}))]^{-1}$.
In the magnetic state $M_{v1}$, the spin polarization $m_2$ develops due to the nonzero one-vacancy component $m_2^1=n_{2\uparrow}^1-n_{2\downarrow}^1\approx 1$ (Fig.~\ref{fig3}(a), right plot). In contrast, the configurations without vacancies and with 
two and more vacancies near Ti do not contribute to the spin-polarization ($m_2^0=m_2^2=m_2^3=m_2^4=0$) which signifies that 
the one-vacancy-configurations are the dominant contribution to the spin polarization of the $e_g$-band in the phase $M_{v1}$.  

For $M_{v2}$ neither the cusp point coordinates $(n_c,c_c)$ nor the extent of the magnetic region are affected by a 
change of the orbital shift $\Delta_v$ between $4t_1$ and $6t_1$.
Therefore we can expect that the the single-vacancy-contributions  do not play a significant role 
in the development of the state $M_{v2}$. This feature is confirmed by the analysis of the band factors $n_{\alpha\sigma}^j$, which shows that the nonzero spin polarization in this case appears
due to the nonzero two-vacancy-component $m_2^2=n_{2\uparrow}^2-n_{2\downarrow}^2\approx 0.1-0.37$ (Fig.~\ref{fig3}(b), right plot).  The corresponding electron filling 
$n_2$ of the $e_g$-band is much higher than
the filling $n_1$ of the itinerant ($\alpha=1$)-band (Fig.~\ref{fig3}(b)) that also points towards a double-exchange Zener mechanism by which two-band magnetic order develops from the local magnetic moments of clusterized vacancies and is stabilized through the kinetic exchange.  

Fig.~\ref{fig4} shows the k-resolved magnon bands calculated from the pole equations of the two-band RPA susceptibility:
$\prod_{\alpha}(1-U\chi_{\alpha}^{+-}(\vec{q},\omega))-J_h^2\prod_{\alpha}\chi_{\alpha}^{+-}(\vec{q},\omega)=0$, where the configurationally averaged Lindhard function is
$\chi_{\alpha }^{+-}(\vec{q},\omega)=(1-c)^4\sum_j \frac{p_j} {N}\sum_{\vec{k}} \frac{n_{\vec{k}+\vec{q},\alpha,\uparrow}^j-n_{\vec{k},\alpha,\downarrow}^j}
{\omega+\Delta_{\vec{k}\vec{q},\alpha}^j}$,
and $\Delta_{\vec{k}\vec{q},\alpha}^j=\lambda_{\alpha,\downarrow}^j(\vec{k})-\lambda_{\alpha,\uparrow}^j(\vec{k}+\vec{q})$.
In both magnetic states $M_{v1}$ (Fig.~\ref{fig4}(a)) and $M_{v2}$ (Fig.~\ref{fig4}(b)), similar features in the magnon spectrum
include the soft Goldstone character of the magnon mode in the $\Gamma$ point for all defect concentrations. In the $M_{v2}$ state, approaching the critical defect concentration $c_c=0.12$, which terminates the magnetic states in the phase diagram of Fig.~\ref{fig2}, leads also to a magnon softening in the vicinity of the $M(\pi,\pi)$ point. 
A similar softening near the $M$ point and close to the $X(\pi,0)$ point is found in the $M_{v1}$-state when the critical defect density $c\approx 0.034$ is approached.
The magnon softening in $M_{v2}$ state around $M(\pi,\pi)$ is accompanied by a dispersion flattening in the range $c\approx 0.15-0.16$. This anomalous behavior reflects the formation of complex 
nearly dispersionless spin-wave states  \cite{moch,lyo,alben}. Fig.~\ref{fig4}(c),(e) and Fig.~\ref{fig4}(d),(f) show the contours of the imaginary parts (intensities) of the static Lindhard functions $\chi_{\alpha}^{+-}(\vec{q},0)$ in the $M_{v1}$ and $M_{v2}$-states. The itinerant component ($\alpha=1$) always 
has a pronounced central peak at the $\Gamma$-point (Fig.~\ref{fig4}(c)) or close to the $\Gamma$-point (Fig.~\ref{fig4}(d)) whereas the localized component ($\alpha=2$) is characterized by high-intensity fragments that develop at nonzero nesting points $(\pm \pi/2,0)$ and $(0,\pm \pi/2)$ (Fig.~\ref{fig4}(e)) and extend to the $M$-points with increasing $c$ (Fig.~\ref{fig4}(f)). The striking difference between the multiband contributions shows that the ferromagnetic order emerges 
with the spin polarization of the itinerant band, whereas the magnetic polarization of the localized band is 
controlled by nonzero wavevectors that introduce complex magnetic oscillations \footnote{Magnetic oscillations in coordinate space will appear due to the coupling/scattering of spin density waves
generated by localized defects or defect clusters in an itinerant ferromagnetic background.}.

In the phase-separated state of the phase $M_{v1}$, the key parameter characterizing the stability is the radius of critical magnetic droplets $R_c$ which can be estimated from the minimization of the contributions of the
Coulomb energy $\varepsilon_C$ and of the droplet surface energy $\varepsilon_s$ in both para- and magnetic states.
Minimization of $\varepsilon_C+\varepsilon_s$ \cite{lorenzana} with respect to $R_c$ results in the expression for the inhomogeneity 
scale: 
$R_c=\left({\sigma}/{u\delta_{fp}^2\xi^{3/2}[\xi^2+(1-\xi)^2]}\right)^{1/3}$,
where $\xi$ is the magnetic phase content, $u=e^2/16\varepsilon$, the charge imbalance between the para- and magnetic states with the densities $n_p$ and $n_f$ is $\delta_{fp}\sim (n_f-n_p)$, and the surface energy per unit surface is given by $\sigma=t_1 c(1-c)^3 \left\{f(m_1,m_2)^2+\frac{t_1}{t_2}  f(m_2,m_1)^2 \right\}$, where $f(x_1,x_2)=(Ux_1+J_H x_2)/t_1$
and $\varepsilon$ is the dielectric constant.
The dependence of $R_c$ on the vacancy concentration $c$ is strongly nonlinear,
due to the competition between the increase of the surface energy $\sigma \sim c$ for larger values $c$ 
on account of pinning and the $c$-driven increase of the 
vacancy-induced
charge-imbalance $\delta_{fp}^2 \sim c^{\zeta}$ with $\zeta\approx 0.5-1$ for $c>c_m$ in the denominator of $R_c$.
In the weak-correlation regime, this leads to a distinct minimum of $R_c$ at $c_m \approx 0.08$, with further
linear enhancement of the droplet radius for $c>c_m$ due to the increase of the magnetic surface energy.    
With $\varepsilon\approx 100$, the characteristic values $R_c$ are in the range of $20a-100a$. This implies
a mesoscopic to microscopic scale of the stable magnetic droplets which is 
by one to two orders of magnitude
larger than the droplet sizes obtained in the scenario of metal-insulator polaron phase separation \cite{nanda} with polaron-type localization \cite{nanda,pavlenko_kopp_prl}, but in agreement with the characteristic scale of inhomogeneities estimated in recent SQUID studies \cite{bert}.

To conclude, we have revealed unconventional magnetic states common for the titanate interfaces which are stabilized above a
finite critical concentration of oxygen defects. 
The magnetic states are characterized 
by an interplay of ferromagnetic order in the itinerant $t_{2g}$ band 
and complex magnetic oscillations
in the orbitally reconstructed $e_g$-band, which at oxide interfaces can be tuned by gate fields.
The magnetic phase diagram includes highly fragmented regions of stable or phase-separated magnetic states emerging 
at nonzero critical defect concentrations, 
mediated by different types of local vacancy configurations.

This work was  supported by the DFG (TRR~80). Grants of computer time from the 
Leibniz-Rechenzentrum M\"unchen through the SuperMUC project pr58pi are thankfully acknowledged.

\end{document}